\documentclass[iop]{emulateapj}

\newcommand{\etal}{et al.}

\newcommand\chandra{{\it Chandra}}
\newcommand\xmm{{\it XMM-Newton\/}}
\newcommand\ro{{\it ROSAT\/}}
\newcommand\fermi{{\it Fermi}}
\newcommand\calv{1RXS J141256.0$+$792204}

\def\simlt{\mathrel{\hbox{\rlap{\hbox{\lower4pt\hbox{$\sim$}}}\hbox{$<$}}}}
\def\simgt{\mathrel{\hbox{\rlap{\hbox{\lower4pt\hbox{$\sim$}}}\hbox{$>$}}}}

\slugcomment{}

\received{2011 March 16}
\accepted{2011 June 10}

\shorttitle{Is Calvera a Gamma-ray Pulsar?}
\shortauthors{Halpern}

\begin{document}

\title{Is Calvera a Gamma-ray Pulsar?}

\author{J. P. Halpern}
\affil{Astronomy Department, Columbia University,
550 West 120th Street, New York, NY 10027-6601, USA;
jules@astro.columbia.edu}

\begin{abstract}

Originally selected as a neutron star (NS) candidate in the
ROSAT All-Sky Survey, \calv\ (``Calvera'') was discovered
to be a 59~ms X-ray pulsar in a pair of \xmm\ observations \citep{zan11}.
Surprisingly, their claimed detection of this pulsar in
\fermi\ $\gamma$-ray data 
requires no period derivative, severely restricting its
dipole magnetic field strength, spin-down luminosity, and distance
to small values.  This implies that the cooling age of Calvera
is much younger than its characteristic spin-down age.  If so, it could be
a mildly recycled pulsar, or the first ``orphaned'' central compact
object (CCO).  Here we show that
the published \fermi\ ephemeris fails to align the pulse phases of the two
X-ray observations with each other,
which indicates that the \fermi\ detection is almost
certainly spurious.  Analysis of additional \fermi\ data
also does not confirm the $\gamma$-ray detection.
This leaves the spin-down rate of Calvera less
constrained, and its place among the families
of NSs uncertain.  It could still be either a normal pulsar,
a mildly recycled pulsar, or an orphaned CCO.

\end{abstract}


 \section {Introduction}

The NS candidate \calv,
dubbed ``Calvera'' \citep{rut08},
was selected from the \ro\ All-Sky Survey, and observed
by \chandra\ \citep{rut08,she09}.  A deep radio pulsar
search showed that it is radio quiet \citep{hes07}.
It was not until a pair of \xmm\ observations was obtained
with high time resolution that \citet{zan11} discovered
59~ms pulsations from Calvera.  The classification
of Calvera among the families of NSs
is not yet understood.  Its X-ray spectrum is characterized as
a blackbody of temperature $\approx 0.2$~keV, or a
hydrogen atmosphere of $T \approx$~0.1 keV \citep{she09,zan11}.
Two-temperature models provide a better fit, and
the surface temperature must be nonuniform because pulsations
are seen.  Calvera's properties distinguish it from
seven isolated NSs \citep[INSs:][]{hab07},
also discovered by \ro, which are slowly rotating ($P = 3-11$~s),
cooler NSs in the solar neighborhood.  X-ray timing and
spectroscopy and kinematic studies of the INSs indicates that
they have strong magnetic fields, $B_s \approx 2 \times 10^{13}$~G,
and are $\approx 10^6$~years old \citep{kap09}.
Calvera is at least twice as hot as the INSs, indicating an age
$\leq 10^5$~yr according to minimal NS cooling
curves \citep{pag04,pag09}. Even though it is at high Galactic latitude
$(\ell,b)=(118^{\circ},+37^{\circ})$, if Calvera is a passively
cooling NS it must be close to its birth place in the disk,
implying a maximum distance of a few hundred parsecs.  Depending on the
X-ray spectral model fitted, the column density is consistent with a
range of values that does not further constrain the distance.
Based on the spectral fits of \citet{zan11}, the bolometric
flux is uncertain by about a factor of 2, and the luminosity is
$L_X \approx 1.7 \times 10^{31}\ d_{300}^2$ erg~s$^{-1}$,
where $d_{300}$ is the distance in units of 300~pc.
Calvera remains radio quiet even after a deeper
search for radio pulsations at 59~ms \citep{zan11}.

Analyzing data from the \fermi\ Large Area Telescope (LAT),
\citet{zan11} claimed that 59~ms pulsations are detected
from \calv\ at $>100$~MeV.  Apart from the marginal significance
of the detection, this result is surprising because
their ephemeris requires
no frequency derivative over the 21 month time span analyzed.
Their effective $2\sigma$ upper limit
is $|\dot f| < 2.6 \times 10^{-16}$ Hz~s$^{-1}$,
implying spin-down power
$\dot E = -4\pi^2I f \dot f < 1.7 \times 10^{32}$ erg~s$^{-1}$,
characteristic age $\tau_c \equiv -f/2\dot f > 1.0 \times 10^9$~yr,
and magnetic field strength
$B_s = 3.2 \times 10^{19}\sqrt{-\dot f f^{-3}}\, < 7.4 \times 10^{9}$~G.
Here we assume a moment of inertia $I = 10^{45}$
g~cm$^2$, $B_s$ is the equatorial surface dipole field,
and frequency $f = 16.89$~Hz.
(We are unable to account for their
quoted upper limit $B_s < 5 \times 10^{10}$~G,
which would allow $\dot f = -1.2 \times 10^{-14}$ Hz~s$^{-1}$.)

These timing parameters imply that \calv\ is not just a
passively cooling NS, but is converting a large fraction
of its meager spin-down power into $\gamma$-rays.  The quoted pulsed
$\gamma$-ray luminosity is
$L_{\gamma} \approx 1.5 \times 10^{32}\ d^2_{300}$ erg~s$^{-1}$
assuming isotropic emission.
All other $\gamma$-ray pulsars have
$\dot E > 2 \times 10^{33}$ erg~s$^{-1}$ \citep{abd10a};
Calvera would be the least energetic $\gamma$-ray pulsar
by an order of magnitude.
In this Letter, we show that the \fermi\
detection of Calvera is almost certainly false.
We then discuss the implications for the nature
of Calvera.

\begin{deluxetable*}{lcccccc}
\tabletypesize{}
\tablewidth{0pt}
\tablecaption{\xmm\ EPIC pn Timing of \calv}
\tablehead{
\colhead{ObsID} & \colhead{Date (UT)} & \colhead{Date (MJD)} &
\colhead{Span (s)} & \colhead{Exp. (s)} &
\colhead{Frequency (Hz)\tablenotemark{a}} & \colhead{$Z_1^2$}
}
\startdata
0601180101 & 2009 Aug 31  & 55,074.30 & 19,911 & 13,941 & 16.8924052(25) & 141.1 \\
0601180201 & 2009 Oct 10  & 55,114.18 & 27,816 & 19,477 & 16.8924041(15) & 201.9
\enddata
\tablenotetext{a}{1 sigma error in parenthesis.}
\label{tab1}
\end{deluxetable*}

\begin{figure}
\begin{center}
\vspace{0.12in}
\includegraphics[width=0.75\linewidth,angle=0]{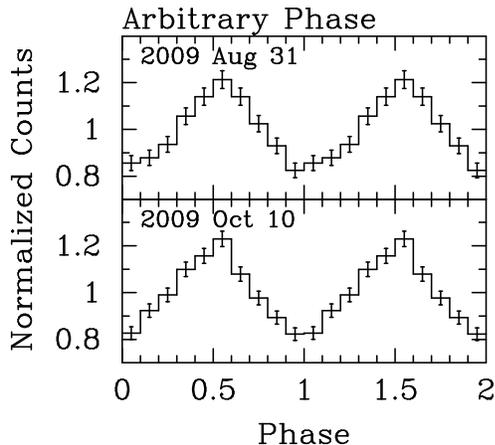}
\vspace{-0.05in}
\end{center}
\caption{
The two \xmm\ pulse profiles of \calv\ in the $0.15-2$~keV band,
each folded on its peak frequency from Table~\ref{tab1}, 
and aligned arbitrarily in phase.  Background has been subtracted
and the counts are normalized to 1.  Two cycles are plotted.
}
\label{fig1}
\end{figure}

\section{XMM-Newton Timing of the Pulsations from \calv}

We reduced the now archival EPIC pn CCD data from the two \xmm\ observations
of Calvera listed in Table~\ref{tab1}. 
These were taken in small window mode with 5.7~ms time
resolution, and were separated by 40 days.
The data were processed with SAS version xmmsas\_20090112\_1802-8.0.0.
We extracted photons in the $0.15-2$~keV band
from a $20^{\prime\prime}$ radius aperture
around the source.  Background was taken
from an adjacent region on the CCD.  We
applied the conversion to Barycentric Dynamical Time
using the precise \chandra\
position of the source from \citet{she09},
R.A. = $14^{\rm h}\,12^{\rm m}\,55.\!^{\rm s}84$,
Decl. = $+79^{\circ}\,22^{\prime}\,03.\!^{\prime\prime}7$ (J2000.0).
The $0.\!^{\prime\prime}6$ position uncertainty is a negligible
source of error on the absolute timing. Table~\ref{tab1} 
lists the peak frequencies of the two \xmm\ observations
derived from a $Z_1^2$ power spectrum
(Rayleigh test; \citealt{str80,buc83}).
We folded each light curve at the peak period.
The pulse profiles are shown in
Figure~\ref{fig1}, where we have aligned them arbitrarily in phase.
The two profiles have a consistent, quasi-sinusoidal
or triangular shape, and a pulsed-fraction of $\approx 18\%$.
The precise agreement between the frequencies
at the $10^{-7}$ fractional level argues
against any orbital motion (but see below), which is 
supported by the absence of an optical counterpart.

As noted by \citet{zan11}, the precision
of the two frequency measurements is insufficient
to join these widely spaced observations coherently and
obtain more precise timing parameters.
Since the measured frequencies agree within their
errors, we can only derive an upper limit on
the spin-down rate.
Using the data in Table~\ref{tab1},
we calculate an upper limit on the frequency derivative
by propagating the errors and dividing the
difference of the frequencies by the time interval
between the observations.  The $2\sigma$ upper
limit is $|\dot f| < 2.0 \times 10^{-12}$ Hz~s$^{-1}$.
The corresponding $2\sigma$ limits on the
spin-down properties are
$\dot E < 1.3 \times 10^{36}$ erg~s$^{-1}$,
$\tau_c > 1.3 \times 10^5$~yr,
and $B_s < 6.5 \times 10^{11}$~G.
These are the limits that we will
conclude are the best currently available.
They are consistent with the \citet{zan11}
results of the same analysis.

\begin{figure}
\begin{center}
\vspace{0.1in}
\includegraphics[width=0.75\linewidth,angle=0]{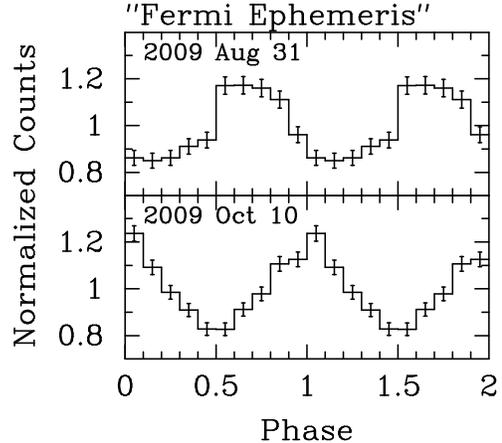}
\vspace{-0.08in}
\end{center}
\caption{
The same data as in Figure~\ref{fig1},
now folded on the supposed \fermi\
ephemeris \citep{zan11} that spans these epochs (see text).
The bin size of the
10 bin light curve is $5.9$~ms, comparable to the CCD frame time
of 5.7~ms. Evidently the ephemeris does not phase-align the X-ray
pulse profiles.
}
\label{fig2}
\end{figure}

Using \fermi\ LAT data, \citet{zan11} claimed to detect
$\gamma$-ray pulsations from \calv\ by searching coherently a
21 month span from 2008 August to 2010 April using
the $Z_1^2$ test.  Although no imaging detection
of a source at this position was reported,
searching both $f$ and $\dot f$ around
the X-ray measured values, they found a peak power
$Z_1^2 = 26.2$ with the following ephemeris:
epoch $t_0 = 55094$ MJD, $f = 16.892401975(2)$~Hz,
$\dot f = -1.2(7) \times 10^{-16}$ Hz~s$^{-1}$.
Since this ephemeris spans
the dates of the \xmm\ observations in 2009 August
and October, it can be used
fold the X-ray photons, and the pulse profiles so
derived should align in phase.
\citet{zan11} did not perform this test.  The result
is shown in Figure~2.  The test fails
because the light curves folded according to the trial ephemeris
are out of phase by 0.35~cycles (21~ms).  As we discuss
below, this is much larger than the maximum $\sim 3$~ms
uncertainty due to the 5.7~ms CCD frame time.
Therefore, we conclude that the ephemeris
represents a noise peak in the power spectrum
of (probably) background photons, not a true detection
of pulsations.

This test does not involve the
phasing of the $\gamma$-ray pulse with respect to the X-ray pulse,
nor is it sensitive to uncertainties on the \fermi\
ephemeris parameters.  This is because the coherent ephemeris
is a {\it phase} ephemeris. It specifies the phase of the pulsar
$\phi(t) = f(t-t_0) + (1/2)\dot f(t-t_0)^2$
at all times during the span of the ephemeris,
and the \xmm\ observations are contained
well within that span.
The quoted \fermi\ uncertainties, $\sigma_{\gamma}(f)=2\times 10^{-9}$~Hz
and $\sigma_{\gamma}(\dot f)=7\times 10^{-17}$ Hz~s$^{-1}$,
can contribute only tiny drifts in relative phase,
$\Delta\phi = \sigma_{\gamma}(f)\,T_X = 0.007$ cycles and
$\Delta\phi = 0.5\,\sigma_{\gamma}(\dot f)\,T_X^2 = 0.0004$ cycles,
respectively, over the $T_X=40$~day interval between the X-ray observations.
Even these are upper limits, as the errors are covariant.

Our analysis does depend on the stability
of EPIC pn timing in small window mode.
Considerable effort has gone into calibrating the relative
and absolute timing of this particular mode
and maintaining the accuracy of the photon
time assignments in the processing chain.
The absolute accuracy of the \xmm\ clock
is better than 0.6~ms \citep{kir04}.
We have carried out extensive investigations
of pulsars using the pn small window mode,
and empirical checks
for consistency show that the absolute times are
at least as accurate as the $\sim 3$~ms
uncertainty due to the 5.7~ms CCD frame time.
We summarize two of these studies here: A coherent
ephemeris for the 237~ms pulsar Geminga
using 10 \xmm\ observations over a span of 7 years
has phase residuals of 2~ms. We have used it to
demonstrate the stable phase relationship
between the X-ray and $\gamma$-ray pulse of Geminga with
{\it AGILE} \citep{pel09} and \fermi\ \citep{abd10b}
to $< 1$~ms.
These agree with earlier results comparing
EGRET and {\it ASCA\/} \citep{jac05}.
In an extensive campaign on the 105~ms pulsar PSR J1852+0040
\citep{hal10}, 16 \xmm\ observations and seven \chandra\
observations (with time resolution 3~ms) 
spanning 4.8 years are fitted by a quadratic
phase ephemeris with rms phase residuals of 3.4~ms.
This demonstrates that \xmm\ and \chandra\ agree
in absolute time to better than 3~ms.

Our analysis does not make use of the EPIC MOS detector
timing data that were obtained simultaneously with the pn.
\citet{zan11} noted that the MOS detector's
timing mode is not well calibrated, and they did
not use it for timing analysis.

\section{Discussion of Previous Fermi Analysis}

We address here reasons why one might
consider the \fermi\ detection of Calvera to be
real despite our negative evidence.
First, \citet{zan11} suggest that actually a more
conservative upper limit on the frequency derivative
from \fermi\ should be allowed, $|\dot f| < 10^{-15}$ Hz~s$^{-1}$.
Still, its effect over the 40 day
interval between X-ray observations
would be negligible, and it would not
change the outcome of our X-ray phase comparison.
It is not clear why they entertain this
possibility, since such a value
would contribute 1.5 extra cycles of rotation over
the 21 month span of their ephemeris.
If $\dot f$ actually turns out to be
$-1 \times 10^{-15}$ Hz~s$^{-1}$, it would be a different
ephemeris from the published one, and the claimed 
detection would be spurious.
For that matter, the frequency
of the pulsar could also turn out to differ by
more than one Fourier bin ($1/T_{\gamma}$, where
$T_{\gamma}$ is the 21 month span of the \fermi\ data)
from the published \fermi\ ephemeris, and still be
consistent with the X-ray measured value.
In this case as well, the claimed $\gamma$-ray
detection is just noise.

Second, it may be argued that, while the fitted
ephemeris corresponds to the average values of
$f$ and $\dot f$ over the time span, there could
be timing noise that smears the pulse, while the
signal is not strong enough to fit such trends
with higher order terms.  Under this hypothesis,
the phase drift between the two
X-ray observations is a manifestation
of timing noise.  We consider this unlikely
because the fitted $\dot f$ is already consistent
with zero.  Any detectable timing noise would
vary the sign of $\dot f$,
which has not been seen in any isolated pulsar
apart from glitch discontinuities.
A phase drift of 0.35 cycles over 40 days would require
an effective $\dot f = -5.9 \times 10^{-14}$ Hz~s$^{-1}$
over this time, almost 500 times the mean value of
$-1.2 \times 10^{-16}$ Hz~s$^{-1}$.
This seems unlikely, as does a glitch that is not also
detected in the $\gamma$-ray timing.  No pulsar
with $\tau_c > 20$~Myr has been observed to glitch
\citep{esp11}.

As an alternative to timing noise, it may be hypothesized
that the X-ray phase shift over 40 days is evidence
of binary motion, i.e, a planetary
companion.  Such an explanation would require 
the orbital period to be much less than the 21 month span
of the \fermi\ ephemeris,
but longer than the durations of the individual \xmm\ pointings,
which are 5.5 hours and 7.7 hours, respectively. 
We consider that $1\,{\rm day}\,<\,P_{\rm orb}\,<\,100\,{\rm days}$
covers the applicable range. 
Under the binary hypothesis, the
projected radius $a_{ns}\,{\rm sin}\,i$
of the NS orbit falls in a narrow range.
The Roemer delay $2\,a_{ns}\,{\rm sin}\,i/c$ must be
$\ge 21$~ms to produce the phase shift,
but $\le 59$~ms so as not to smear out the supposed
$\gamma$-ray pulsations.  Combining these requirements,
for $m_{ns} = 1.4\,M_{\sun}$ we find that
$$20\,M_{\earth} \ < 
\left({P_{\rm orb} \over 100\,{\rm days}}\right)^{2/3} m_p\ {\rm sin}\,i < \ 
60\,M_{\earth}.$$
This corresponds to a minimum planet mass of
$m_p = 20/{\rm sin}\,i$ Earth masses, and
a maximum of $4/{\rm sin}\,i$ Jupiter masses, the latter
for $P_{\rm orb} = 1$ day.  More likely
the $\gamma$-ray ephemeris is spurious,
and the 0.35 cycle phase shift is
a random number, not evidence of a planet.

Other aspects of the \citet{zan11} analysis
are unusual.  By their own description, the $\gamma$-ray signal is
marginal, with $Z_1^2 = 26.2$, and its significance
depends on the number of independent trials in the search.
The trials can be assessed using the X-ray uncertainties
on the timing parameters in Section~2,
$\sigma_X(f) = 1.5 \times 10^{-6}$~Hz and
$\sigma_X(\dot f) = 1 \times 10^{-12}$ Hz~s$^{-1}$,
and the $T_{\gamma}=21$ month
span of the \fermi\ data. The number of independent 
trials in the two-dimensional \fermi\ search should be at least
$4\sigma_X(f)\,T_{\gamma} \approx 330$ for frequency, and
$\sigma_X(\dot f)\,T_{\gamma}^2 \approx 3000$
for frequency derivative.  These represent a search of
independent frequencies in the $\pm 2\sigma$ interval around
the \xmm\ measured $f$, and independent frequency derivatives ranging
from zero to the $-2\sigma$ limit, i.e., only negative $\dot f$.
If so, the expected number of noise peaks of power $Z_1^2 \geq 26.2$,
obtained by multiplying the $1\times 10^6$ trials by
the single-trial probability, $e^{-26.2/2}=2\times 10^{-6}$,
is of order unity.  The oversampling by a factor of 10 that was
performed further reduces the statistical significance.
The crux of their argument
must be that, since the value of $\dot f$ in the discovered
signal is consistent with zero, almost no trials in $\dot f$ were
needed to find it.  Only this would allow 
that the chance probability of the result is $\sim 7 \times 10^{-4}$.
But they do not display the power
spectrum for the complete search. Instead, they say
that they did not find false detections over
a range of parameters much wider than the X-ray uncertainties.
It is not stated what they consider a false detection,
and we are not shown the values of the highest peaks
in the extended search.  Therefore, we don't know what
to make of this argument.
Finally, the absence of the $\gamma$-ray
source in a spatial image is also worrisome, especially since
the position is at high Galactic latitude with minimal
confusing diffuse background.  So there is no supporting
evidence of a source at this position.

\begin{figure}
\begin{center}
\includegraphics[width=0.7\linewidth,angle=270]{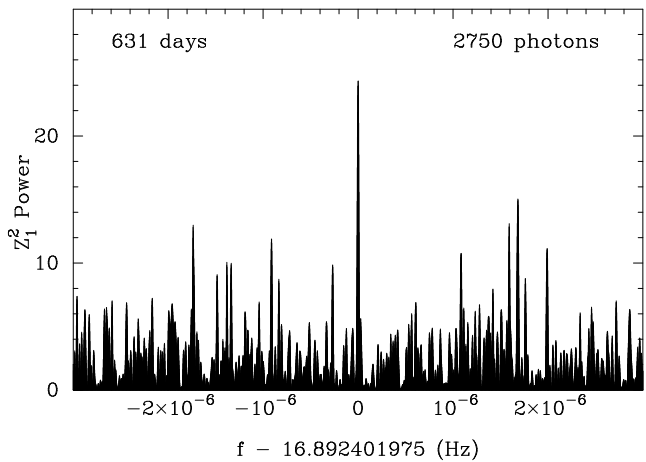}
\end{center}
\caption{
Power spectrum of \fermi\ data covering the time span
analyzed by \citet{zan11}, 2008 August 4  -- 2010 April 27.
The peak power $Z_1^2 = 24.4$
is similar to their value of 26.2.
The two-dimensional search in $f$ and $\dot f$
was projected onto one dimension in this plot.
}
\label{fig3}
\end{figure}

\section{Analysis of New Fermi Data}

For completeness, we extracted and reduced \fermi\ data using
the same event filtering and methods described by \citet{zan11}.
We first extracted the identical 631 day time span, 2008 August 4 --
2010 April 27, recovering 2750 photons, similar to their 2518 photons.
A $Z_1^2$ search covering their $\pm 3\sigma$ range of
\fermi\ frequency derivative recovers the candidate peak at
$f = 16.892401976$, as shown in Figure~\ref{fig3}.
The peak power, $Z_1^2 = 24.4$, is consistent with
their value of 26.2.

\begin{figure}
\begin{center}
\includegraphics[width=0.7\linewidth,angle=270]{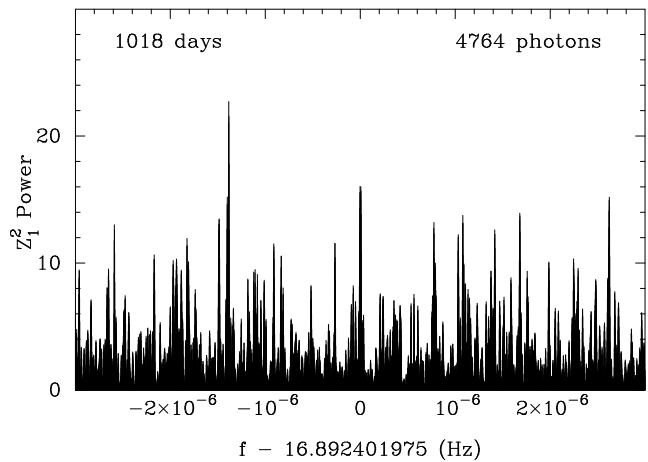}
\end{center}
\caption{
Power spectrum of \fermi\ data covering an extended time span,
2008 August 4 -- 2011 May 19, showing that the peak in
Figure~\ref{fig3} is no longer significant.
}
\label{fig4}
\end{figure}

We then applied the same method to the full data set now available,
comprising 4764 photons collected up to 2011 May 19 (33 months).
This increases
the number of photons by 73\%.  A search of the same
ephemeris parameters does not yield increased power at the claimed
frequency.  Rather, the peak previously seen is reduced to $Z_1^2 = 16.0$
(Figure~\ref{fig4}), indicating that it was just noise.
We regard this result as direct support of our
inference that the supposed \fermi\ detection was not real.
A wider search of thousands of trials in $\dot f$ is not
meaningful, for the reasons discussed above.

\section{Conclusions}

In summary, the incorrect phasing of the
X-ray observations of Calvera
using the claimed \fermi\ ephemeris
led us to conclude that the $\gamma$-ray
detection is probably spurious.  Then,
extending the \fermi\ analysis from 21 to 33 months
rendered the candidate signal insignificant.
Here we comment
on the implications for the nature of Calvera
of the less restrictive $2\sigma$ limit
$|\dot f| < 2.0 \times 10^{-12}$ Hz~s$^{-1}$
from X-ray timing.
The corresponding limits on spin-down properties,
$\dot E < 1.3 \times 10^{36}$ erg~s$^{-1}$,
$\tau_c > 1.3 \times 10^5$~yr,
and $B_s < 6.5 \times 10^{11}$~G,
allow three scenarios.
First, Calvera could be an ordinary pulsar with magnetic
field strength $1.5\sigma$ or more below the mean of the
birth distribution \citep{fau06}.  Second,
it could be a mildly recycled pulsar, formerly in a
binary system with a high-mass companion that has since
undergone a supernova explosion. In this case, its
magnetic field strength would be intermediate between those
of fully recycled (millisecond) pulsars, and ordinary pulsars.
Third, it could be an ``orphaned CCO'' as we describe next.
The second and third scenarios were also addressed by \citet{zan11}.

The class of central compact objects (CCOs) in supernova
remnants comprises $\approx 10$ NSs, three
of which are detected pulsars with $P = 0.1-0.4$~s.
See \citet{hal10,hal11} and \citet{got10}
for observations and overview of related theory.
The CCO pulsars
have weak dipole fields, in the range $10^{10}-10^{11}$~G,
and negligible spin-down power in comparison with their
bolometric X-ray luminosities of $10^{33}-10^{34}$ erg~s$^{-1}$.
These $10^3-10^4$~yr old NSs
must represent a significant fraction of NS births.  After
their host supernova remnants dissipate, orphaned CCOs
will remain in the region of $(P,\dot P)$ space where they
were born, which is also where the supposed mildly
recycled pulsars are found \citep{bel10}.
An orphaned CCO would be distinguished from a single,
recycled pulsar by its residual
thermal X-ray luminosity.  An orphaned CCO could be recognized as
a thermal X-ray source, depending on its distance,
while it is up to $10^5-10^6$~yr old
(not the characteristic age, which is
orders of magnitude larger than the real age of a CCO).
The known CCOs have X-ray temperatures in the range $0.2-0.4$~keV.
Calvera, being cooler than this and less luminous than a young
CCO by a order of magnitude or more, could be an evolved stage
of a passively cooling CCO.  It is possible, therefore, that
\calv\ is the first orphaned CCO to be recognized.

These scenarios can be distinguished
by the spin-down rate of the pulsar.
If $B_s = 6 \times 10^{11}$~G and $\tau_c = 1.5 \times 10^5$~yr,
Calvera is probably an ordinary pulsar.
If $B_s = 1 \times 10^{11}$~G and $\tau_c = 4 \times 10^6$~yr,
it could be an orphaned CCO younger than $\tau_c$,
or a spin-powered, mildly recycled pulsar.
If $B_s = 1 \times 10^{10}$~G and $\dot E = 3 \times 10^{32}$ erg~s$^{-1}$,
its spin-down power is probably insufficient
to heat its surface, and
an orphaned CCO would be required instead of a mildly
recycled pulsar to explain its X-ray luminosity and temperature.
Even though Calvera is not (yet) detected in
$\gamma$-rays, an X-ray timing study is
straightforward, and should detect its spin-down
in $\approx 1$~yr even if its magnetic field strength is
only $\approx 10^{10}$~G.

\acknowledgements

We thank Eric Gotthelf for discussions and assistance with
the data. This investigation is based on observations obtained with \xmm,
an ESA science mission with instruments and contributions directly funded by
ESA Member States and NASA.

\end{document}